\documentstyle[prl,aps,preprint]{revtex}

\begin{document}
\tighten

\title{EXACT MATCHING CONDITION FOR\\
MATRIX ELEMENTS IN LATTICE AND ${\rm \overline{MS}}$ SCHEMES \thanks
{This work is supported in part by funds provided by the U.S.
Department of Energy (D.O.E.) under cooperative
agreement \#DF-FC02-94ER40818.}}

\author{Xiangdong Ji }

\address{Center for Theoretical Physics \\
Laboratory for Nuclear Science \\
and Department of Physics \\
Massachusetts Institute of Technology \\
Cambridge, Massachusetts 02139 \\
{~}}

\date{MIT-CTP-2447 \hskip 1in  hep-lat/xxxxxxx \hskip 1in June 1995}

\maketitle

\begin{abstract}
The exact matching condition is given for hadron matrix elements
calculated in any two different schemes, in particular, in the lattice
and dimensional regularization, (modified) minimal subtraction
$\overline{\rm MS}$ schemes. The result provides insight
into and permits to go beyond
Lepage and Mackenzie's mean field theory of
removing tadpole contributions in
lattice operators.

\end{abstract}

\pacs{xxxxxx}

\narrowtext
Presently, lattice QCD provides the unique
method with controlled approximation to compute
hadron properties directly from the
QCD lagrangian. In the last few years, a number of groups
have calculated on the lattice an impressive list of hadron matrix
elements, ranging from the axial and scalar charges of the
nucleon to lower-order moments of deep-inelastic structure
functions \cite{MS,GH,LIU}. Note, however, that
most of the hadron matrix elements are not directly physical
observables. In field theory, apart from the S-matrix,
physical observables are related to symmetry generators
of the lagrangian, such as the vector and axial-vector currents
or hadron masses. Nonetheless, hadron matrix
elements are useful intermediate quantities to express
physical observables. Being intermediate, they often depend
on specific definitions in particular context. Or in
field theory jargon, they are scheme-dependent.
Since schemes are generally
introduced to eliminate ultraviolet divergences in composite
operators, the scheme dependence of a matrix element is in fact
perturbative in asymptotically-free QCD.

Understanding scheme dependence has important practical
values. In calculating hadron matrix elements on a lattice,
one is automatically limited to the lattice scheme. On the
other hand, hadron matrix elements entering physical
cross sections are often defined in connection with perturbation
theory. The best scheme for doing perturbation theory
is not the lattice QCD, because the lattice has complicated
Feynman rules and accommodates only Euclidean
Green's functions.
The most popular scheme for perturbative calculations
is the dimensional regularization introduced by t' Hooft
and Veltman more than two decades ago, followed by the (modified)
minimal subtraction (${\rm \overline{MS}}$).

A popular practice currently adopted in the literature for matching
the matrix elements in the lattice and ${\rm \overline{MS}}$ schemes
goes like this \cite{MS,KP}. Consider, for instance, a
quark operator $O$. First, the one-loop matrix element of $O$ in a
single quark state $|k\rangle $ is calculated on the lattice,
\begin{equation}
       \langle k |O| k \rangle_{\rm latt}
    = A\left(1 + {g_0(a)^2\over {4\pi^2}}(\gamma \ln a^2p^2 + c)\right) \ ,
\end{equation}
and in the ${\rm \overline{MS}}$ scheme,
\begin{equation}
       \langle k |O| k \rangle_{\overline {\rm MS}}
    = A\left(1 + {g(\mu)^2\over {4\pi^2}}(\gamma \ln p^2/\mu^2 + c')\right) \ ,
\end{equation}
where $p^2$ is an infrared cut-off.
Second, the hadron matrix elements in the state
$|P\rangle$ in the two schemes are assumed to have the relation,
\begin{equation}
     { \langle P |O| P\rangle_{\rm latt}\over \langle P |O|
       P\rangle_{\overline{\rm MS}}}
    ={ \langle k |O| k \rangle_{\rm latt}\over \langle k |O| k
     \rangle_{\overline{\rm MS}}}
    = 1 + {g^2\over {4\pi^2}}(\gamma \ln a^2\mu^2 + c-c') \ .
\end{equation}
In the second equality, an expansion in $g^2$ is made.
In order to cancel the infrared
cutoff $p^2$, one has to identity $g_0(a)$ and $g(\mu)$.
However, after the cancellation, one does not know which $g$
should be used in the above equation. Due to the large tad-pole
contributions,
the difference between coupling constants in the two schemes
is significant in the same momentum region. Furthermore, it is not
clear how to generalize the above relation to multi-loops.

As indicated early, the difference between the matrix elements
in any two schemes arises from the ultraviolet region.
Thus there must exist an all-order perturbative
relation between the matrix elements calculated in the
two schemes. In this note, I shall derive such a
relation using renormalization group arguments.

Consider a physics observable $M(Q^2)$ where $Q^2$
is a hard momentum scale of a physical process. $M(Q^2)$ could be
a moment of some deep-inelastic structure function. According
to the operator-product expansion, we write,
\begin{equation}
     M(Q^2) = C(Q^2/\mu^2, \alpha_s(\mu^2))A(\mu^2) \ .
\end{equation}
where $C$ is the coefficient function, calculable in perturbative
theory. $C$ is clearly scheme-dependent, and let us
assume a scheme has been chosen for its calculation. $A$ is a
soft hadron matrix element defined in the same scheme.
The $\mu$ dependence of
$A$ satisfies the renormalization group equation,
\begin{equation}
        {dA(\mu)\over d\mu} + \gamma(\mu) A(\mu) = 0 \ ,
\end{equation}
where $\gamma$ is the anomalous dimension of the operator
in the scheme.
The renormalization scale dependence
must cancel between $C$ and $A$, leaving $M$ $\mu$-independent.
Therefore $C$ satisfies the
renormalization group equation,
\begin{equation}
          {dC\over d\mu} - \gamma(\mu) C = 0 \ .
\end{equation}
The above equation can be integrated formally to yield,
\begin{equation}
          C(Q^2/\mu^2, \alpha_s(\mu^2))
         = C(1, \alpha_s(Q^2))\exp\left(-\int^{g(Q)}_{g({\mu})}
{\gamma\over \beta} dg\right) \ ,
\end{equation}
where the QCD $\beta$-function is also scheme-dependent.
 From the above formulas and the fact that $M$ is scheme-independent,
we derive a relation between the
matrix elements in two different schemes,
\begin{equation}
    C(1, \alpha_s(Q^2))\exp\left(-\int^{g(Q)}_{g({\mu})}
  {\gamma\over \beta} dg\right) A(\mu)
 = \tilde C(1, \tilde \alpha_s(Q^2))
   \exp\left(-\int^{\tilde g(Q)}_{\tilde
g(\tilde \mu)}{\tilde  \gamma\over
\tilde \beta} dg\right) \tilde A(\tilde \mu) \ .
\end{equation}
It seems that one has to know the coefficient
functions in the two schemes before hand.
However, this is unnecessary
if one takes $Q^2 \rightarrow \infty$ and uses
asymptotic freedom. In the limit, the coupling constants
in different schemes approach to zero at the same rate
and $C(1,0)$ becomes scheme-independent! Thus
we have,
\begin{equation}
    \exp\left(-\int^0_{g(\mu)}
  {\gamma\over \beta} dg\right) A(\mu)
 =    \exp\left(-\int^{0}_{\tilde
g{(\tilde \mu)}}{\tilde  \gamma\over
\tilde  \beta} dg\right) \tilde A(\tilde \mu) \ .
\end{equation}
This is the desired equation, which depends on the
$\beta$-functions and anomalous dimensions of the operator
in two different schemes. The matrix elements
defined in this equation
satisfy the renormalization group equations in
both schemes.

To specialize to the lattice and $\overline{\rm MS}$ schemes,
let me make the following definitions,
\begin{equation}
    O_B({1\over \epsilon}) = Z(\mu)O_{\overline{\rm MS}}(\mu); \  \  \
    \gamma_{\overline{\rm MS}} = {\mu\over Z(\mu)} {dZ(\mu)\over d\mu} \ ,
\end{equation}
where $O_B(1/\epsilon)$ and $O_{\overline{\rm MS}}$ are the
bare and renormalized operators in the $\overline{\rm MS}$
scheme, and
\begin{equation}
 O_{\rm latt}(a) = Z(a) O_{\rm R~latt}; \  \  \
  \gamma_{\rm latt} = - {a\over Z(a)}{dZ(a)\over da} \ ,
\end{equation}
where $O_{\rm latt}(a)$ and $O_{\rm R~latt}$ are the bare and
renormalized operators in the lattice scheme.
Define further a renormalization relation between the bare
lattice and renormalized $\overline{\rm MS}$ operators,
\begin{equation}
       O_{\overline{\rm MS}}(\mu) = Z(\alpha_s(a),
         \alpha_s(\mu)) O_{\rm latt}(a)  \ .
\end{equation}
Then the renormalization constant is,
\begin{equation}
     Z(\alpha_s(a), \alpha_s(\mu))=  \exp{\left(\int^{g(a)}_0{\gamma_{\rm
latt}\over
\beta_{\rm latt}}dg\right)}
      \exp{\left(-\int^{g(\mu)}_0 {\gamma_{\overline{\rm MS}}\over
\beta_{\overline{\rm MS}}}
   dg\right)}   \ .
\label{z}
\end{equation}
The corresponding relation between the matrix elements
in the two schemes is $A_{\overline{\rm MS}}(\mu)
= Z(\alpha_s(a), \alpha_s(\mu))A_{\rm latt}(a)$.

In practical applications, one uses a
perturbative expansion of Eq. (\ref{z}). To next
to the leading order,
\begin{eqnarray}
       \gamma &=& \gamma_0({\alpha_s\over 4\pi})
     +\gamma_1({\alpha_s\over 4\pi})^2 + ... \ , \nonumber \\
     \beta &=& -g\left (\beta_0({\alpha_s\over 4\pi})
     +\beta_1({\alpha_s\over 4\pi})^2 + ...\right) \ .
\end{eqnarray}
Substituting the above into Eq. (\ref{z}), I find,
\begin{equation}
     Z(\alpha_s(a), \alpha_s(\mu))
    = \left({\alpha_s(\mu)\over \alpha_s(a)}\right)^{\gamma_0/2\beta_0}
       \left({1+ (\alpha_s(\mu)/ 8\pi)({\gamma_1(\mu)\over
      \beta_0}-{\gamma_0\beta_1\over \beta_0^2}) + ...\over 1+
     (\alpha_s(a)/8\pi)({\gamma_1(a)\over
      \beta_0}-{\gamma_0\beta_1\over \beta_0^2}) +...}\right) \ ,
\label{y}
\end{equation}
where I have used the fact that the first-two-loop $\beta$ functions
and the one-loop anomalous dimensions are scheme-independent.
$\gamma_1$ is the two-loop anomalous dimension and is
scheme-dependent. The constants $c$ and $c'$ in Eqs.
(1) and (2) do not enter in the above relation directly.
However, they contribute through the two-loop
anomalous dimensions.

The first factor in Eq. (\ref{y}) sums over all the leading
logarithmic contributions. At the momentum
scales associated with current lattice calculations,
the strong coupling constant in $\overline{\rm MS}$
is nearly a factor 2 larger than the bare lattice
coupling because of the large tadpole contributions. Thus
if $\gamma_0/2\beta_0$ is on
the order of unity, the first factor in Eq. (15)
is large. However, this is not the only source of large
tadpoles contributions. The lattice operators,
defined using the gauge link $U_\mu=\exp(ig_0aA_\mu)$,
where $A_\mu$ is the gauge potential,
also contain large tadpole effects,
reflected by the large constant $c$
in Eq. (1). This large constant induces
a large two-loop anomalous dimension $\gamma_1(a)$.
As a consequence, the lattice perturbation series in the
denominator of Eq. (\ref{y}) converges
sluggishly and must be improved with
some resummation procedure.

In Ref. \cite{LM}, Lepage and Mackenzie
suggested to replace the bare lattice coupling $\alpha_s(a)$
in a perturbative expansion by a renormalized coupling,
such as the $\overline {\rm  MS}$ coupling, $\alpha_s(\mu)$.
Such improvement procedure eliminates large coefficients in the
expansion and speed up the convergence of the
series. However, to completely cancel the tadpole
contributions, one must multiply
the series with an expansion of
some power of $\alpha_s(a)/\alpha_s(\mu)$.
The appropriate power is determined by
the anomalous dimension of the operator, which in turn
depends on a particular construction of the
operator on the lattice. The residual
factor of $\alpha_s(a)/\alpha_s(\mu)$ in Eq. (\ref{y})
represents the intrinsic tadpole effect of
the lattice operator, accounting for
the major part of the renormalization factor.
For the same continuum operator, one
can construct many different versions of lattice operators,
which are affected by tadpole contributions differently.

In general, the lattice two-loop anomalous dimension
$\gamma_1(a)$ can give a good indication as to
what power of $\alpha_s(a)$ is needed to cancel
the tadpole effects in the lattice series.
Lepage and Mackenzie's mean field theory
provides an empirical method to determine
the residual power of $\alpha_s(a)/\alpha_s(\mu)$
in the renormalization factor. Consider a
lattice operator
constructed with $k$ powers of gauge links. Then,
according to Lepage and Mackenzie,
the tadpole effects can be
accounted for by multiplying the lattice operator
with a factor of $u_0^{-k}$. Through studying
the effect on the Wilson action,
Lepage and Mackenzie determined
$u_0 = (\alpha_s(a)/\alpha_s(\mu))^{1/4}$. Thus,
the mean field theory predicts
$Z\sim (\alpha_s(a)/\alpha_s(\mu))^{-k/4}$
apart from a well-behaved perturbation series.

However, Eq. (\ref{y}) allows one to go beyond
the mean field theory with the
explicit dependence on perturbative corrections to high
orders. With more powerful methods becoming
available for computing lattice perturbation series,
this exact formula for the renormalization factor
is quite valuable. The formula also corrects
the mean field prediction when the latter fails. An extreme
case is the unit operator
constructed as $UU^{-1}$ on the lattice, for
which a naive application of the
mean field theory fails completely.

To summarize, I have derived an exact relation between
the matrix elements in lattice and
$\overline{\rm MS}$ schemes. The relation is used to
discuss Lepage and Mackenzie's mean field theory
for correcting the tadpole effects in lattice operators.
The relation is useful for comparing
the hadron matrix elements calculated in the lattice
and measured, for instance, in deep-inelastic
scattering.

\acknowledgments
I wish to thank W. Lin, B. Schrieber,
U. Wiese, and in particular, J. Negele for
useful discussions.

\end{document}